# Thermodynamic Stability of Medium with an Alternating-Sign Pressure

**V.I. Laptev**

*Russian New University, Radio St. 22, Moscow 105005, Russia*

Correspondence should be addressed to Viktor I. Laptev, viktor.laptev@yahoo.com

The thermodynamic medium is treated as homogeneous phase equilibrium with a special feature: the pressure of one phase is positive and the pressure of the other is negative. From this point of view such medium is neither mixture, usual solution nor gel (or sol as a whole) regardless its components (solvent and dissolved substances). To base this statement the theoretical investigation of the conditions of equilibrium and stability of the medium with sign-alternative pressure is carried out under using the thermodynamic laws and the Gibbs" equilibrium criterion. Thermal radiation and condensate are claimers on the medium with alternating-sign pressure.

## 1. INTRODUCTION

A complete description of the thermodynamic state of a medium with the constant mass and without chemical interactions is given by a relationship between the internal energy U, entropy S and volume V [1]. They have the positive values. In thermodynamic S and V conjugate with a temperature T and a pressure p. There are thermodynamic systems with a positive and negative temperature [2]. It is unknown any systems with an alternating-sign pressure. This paper describes thermodynamic stability of the still unknown systems with opposite sign pressure.

## 2. KNOW-HOW

The paper reports the conditions of equilibrium and segregation stability of a medium consisting of a positive pressure phase and a negative pressure one. It is shown that this unusual state of the medium with the negative pressure and positive temperature is not metastable and coexists without segregation in equilibrium with an usual medium state where the pressure and temperature are positive.

## 3. BASIC THERMODYNAMIC DIFINITIONS

The considered medium is a matter as a homogeneous or a heterogeneous thermodynamic system. We suppose that the homogeneous medium can exist as a combination of two phases without segregation. This state is called homogeneous phase equilibrium. In this paper the properties of multiphase homogeneous medium are described.

3.1. Using Model. An initial matter is considered as a homogeneous medium with a constant mass and without chemical interactions.

3.2. U, S, V-Space. The surface $\varphi\,(U, S, V) = 0$ corresponds to the all equilibrium states of the medium. It has a tangency with a plane which is an algebraic surface $\varphi_1\,(U, S, V) = 0$ of the first order in orthogonal coordinates. The coordinates of the tangent points are given by the equation

$$U = TS - pV \qquad (1)$$

after superposition of tangent surfaces at the origin of coordinates. The mathematical procedure supposes using absolute values of the internal energy U and entropy S. They are positive.

Pressure p and temperature T of the medium are determined by a tangent plane to the surface $\varphi\,(U, S, V) = 0$. It is an algebraic surface $\varphi_1\,(U, S, V) = 0$ of the first order in orthogonal coordinates, its slope to the surface $\varphi=0$ determines pressure p and temperature T by the equations

$$p = -(\partial U/\partial V)_S, \qquad (2)$$

$$T = (\partial U/\partial S)_V. \qquad (3)$$

The index of the derivative points out a value which is supposed to be constant under differentiation [1, 3 - 5].

3.3. Usual or Unusual System. The signs of the values U, S, and V are constant. If the internal energy U monotonically und unrestrictably increases with entropy S and volume V, this thermodynamic system is called usual thermodynamic system, temperature and pressure are considered as positive values [5].

According to (1) - (3), p or T has alternating-sign if internal energy U of a system has extremum. Media with finite internal energy are known as unusual thermodynamic systems [5]. The nuclear spin energy of a solid can be such an example because it has upper limes. The temperature of such spin system is alternating-sign [2].

We consider the wanted medium as an unusual system because the internal energy limit is the primary reason for its alternating-sign pressure. The properties of system with the finite internal energy are specific.

3.4. Alternating-sign Pressure and Temperature. The spin temperature of the crystal is alternating-sign; the crystal temperature and pressure are positive only [2]. Unfortunately, the spin pressure sign is not found in those works. It can be found with help of the (1), (2). Obviously its right half is positive at the negative temperature if the pressure is negative. Therefore spin energy can be a reason of the evenly stretch of solid which is compensated by electromagnetic attraction of spin carries.

The effect of the negative pressure is known as a cavitation in fluids and a homogeneous tensile deformation of a solid. The mechanics knows a hypothetical Chaplygin gas with a negative pressure. Cosmology operates with a category of negative pressure for a homogeneous isotropic medium. A cosmologic medium state with a zero pressure is called a "dust"-like one. However, we don't know an example of the medium or phase equilibrium with a sign-alternate pressure.

## 4. THEORY

The first law of thermodynamics for systems with finite energy is unchanged as a law about entropy existing and increasing [5]. We pay a particular attention that from the (1) it follows that the negative T is possible only at the negative pressure p. It should be clear, that this statement does not allow making a so-called reversible statement: it is not correct to think that the negative pressures are possible only at negative temperatures. In this work the alternating-sign pressure at the positive temperature is studied.

4.1. Gibbs' Criterion. The surface $\varphi(U, S, V) = 0$ corresponds to the all equilibrium states of the medium. In algebra the general condition of thermodynamic equilibrium for usual and unusual phases is as follows: the virtual displacements of total entropy $\delta S^* = 0$ under constant $U^*$ and fixed V [1, 3 - 5]. For example, the isolated system from two phases 1 and 2 with ratios $S^*=S_1+S_2$ and $U^*=U_1+U_2$, $T\delta S=\delta U+p\delta V$ and $\delta U_1+\delta U_2=0$ is described by the equation

$$(1/T_1 - 1/T_2)\delta U_1 + (p_1/T_1)\delta V_1 + (p_2/T_2)\delta V_2 = 0 \quad (4)$$

with phase indices only. For any values of variations $\delta U_1$ or $\delta U_2$ we find preliminary equilibrium conditions for the usual and unusual media:

$$1/T_1 - 1/T_2 = 0.$$

If $\delta V_1 = \pm \delta V_2$ from (4) we have

$$(p_1/T_1 \pm p_2/T_2)\delta V_{1,2} = 0$$

for any values of variations $\delta V_1$ or $\delta V_2$.

4.2. Gibbs' Equilibrium Condition for Usual Medium. It is known [3-5], that if $\delta V_1 = -\delta V_2$ the solution of (4) is

$$T_1 = T_2, \qquad p_1 = p_2,$$

In geometry these usual states 1 and 2 are identified by two tangent points of the plane to the surface $\varphi(U, S, V) = 0$ [1]. Phase temperature and pressure change not the sign; such medium is usual thermodynamic system.

4.3. Gibbs' Equilibrium Condition for Unusual Medium. According to (1), the negative pressure is possible only under the negative pressure. Nevertheless, the reverse speculation is not possible: one cannot state that the negative pressure exists under negative temperatures only. So, in the equilibrium thermodynamic system the sign-alternative pressure is possible under positive temperature. For example, here we have found that the isolated system with $\delta V_1 = \delta V_2$ has a solution of (4):

$$T_1 = T_2, \qquad p_1 = -p_2. \quad (5)$$

Geometrically it is presented by two tangent planes to the surface $\varphi=0$ with a total slope to the surface $U = \text{const}$ is $180^0$. Phase temperature and pressure change the sign; such medium is unusual thermodynamic system.

4.4. Medium Segregation Stability. We don't know any analysis of the stability of the sign-alternative pressure medium. Let us examine their stability related to the initial medium using the Gibbs' stability criterion [3-5]. The initial medium with constant mass is homogeneous and isotropic. Indexing its entropy $S_0 = S_1 + S_2$ we rewrite (1) as

$$U_0 - T_0 S_0 + p_0 V_0 = 0. \quad (6)$$

We have to find out which of signs =, > or < appears between the internal energy $U_0$ of the initial medium and the sum $U_1+U_2$ of the phase energies under the sign variation of the pressures $p_1$, $p_2$ and the pressure $p_0$ of the initial medium.

4.5. Segregation Stability of the Usual Medium. Let us consider the decay of the initial medium with formation of phases 1 and 2. If $V_1 + V_2 = V_0$ and $p_1 = p_2$, expressions (1), (6) may be rewritten as

$$U_0 - (p_{1,2} - p_0)V_0 = U_1 + U_2. \quad (7)$$

Then we examine the phase equilibrium stability relating the initial medium under variation of the pressure sign in (7) for three cases. The signs of $p_1$ and $p_2$ are identical.

Scenario 1i. If $p_0 = p_1 = p_2$, instead of (7) we have

$$U_0 = U_1 + U_2.$$

Sign = here means that the initial medium and the phase can be in equilibrium under segregation. To visualize our description we denote the internal energy as $\bar{U}$ when the pressure is negative. Then the sign = in the expression

$$\bar{U}_0 = \bar{U}_1 + \bar{U}_2$$

means that the initial medium and phases can be in equilibrium under negative pressure.

<u>Scenario 2i.</u> If $p_0 = 0$ and $p_1 = p_2 > 0$, instead of (7) we have

$$U_0 > U_1 + U_2.$$

The sign > means that the initial medium is unstable relating to the equilibrium of new phases with a positive pressure under their segregation.

<u>Scenario 3i.</u> If $p_0 = 0$ and $p_1 = p_2 < 0$, instead of (5) one has

$$U_0 < \bar{U}_1 + \bar{U}_2.$$

The sign < means that under segregation of 1 and 2 with negative pressure their equilibrium is unstable relating to the initial medium with a zero pressure.

The pressure $p_1$ and $p_2$ in scenario 1i - 3i have identical signs. Then the phases 1 and 2 are usual media.

4.6. Segregation Stability of an Unusual Medium. Now we ascribe the positive pressure to the phase 1 and the negative pressure to the phase 2. In the geometry of the U, S, V – space it means the intercept of two planes with an unique tangency with the surface φ=0. The interception line is an isotherm of the plane V=const. Then the equilibrium condition (5) is physically valid under absence of the phase segregation and $V_1=V_2=V_0$, where $V_0$ is a volume of the medium. In this case, according to (1), (5), (6),

$$U_0 + p_0 V_0 = U_1 + U_2. \qquad (8)$$

Now we again investigate the phase equilibrium condition relating to the initial medium under variation of the pressure of initial medium $p_0$ in three cases below.

<u>Scenario 1j.</u> When $p_0 = 0$, instead of (8) we have

$$U_0 = U_1 + \bar{U}_2.$$

The sign = means that under $p_0 = 0$ the phases 1 and 2 may be in equilibrium with the initial medium without segregation.

<u>Scenario 2j.</u> If $p_0 > 0$, instead of (8) we have

$$U_0 < U_1 + \bar{U}_2.$$

The sign < means that the phases 1 and 2 are unstable relating to the initial medium with a positive pressure without segregation.

<u>Scenario 3j.</u> When $p_0 < 0$, instead of (8) we have

$$\bar{U}_0 > U_1 + \bar{U}_2.$$

The sign > means that the initial medium with a negative pressure is unstable relating to the equilibrium of new phases without segregation.

The pressure $p_1$ and $p_2$ in scenarios j have opposite signs. Then phases 1 and 2 combine unusual media.

4.7. Compensation of Usual and Unusual Media. I-j scenarios have a different energy balance for the initial medium and new phases and suppose compensation. Assume that the initial medium spontaneously decays according to the scenario 3j. Corresponding to (5), the uncompensated energy is equal to the difference

$$\Delta U = \bar{U}_0 - (U_1 + \bar{U}_2) = -p_0 V_0 > 0.$$

When the initial medium spontaneously decays according to the scenario 2i, we have

$$\Delta U = U_0 - (U_1 + U_2) = p_{1,2} V_0 > 0.$$

The uncompensated energy $\Delta U$ can be dissipated in space.

The zero difference $\Delta U = 0$ corresponds to the scenario 1i or 1j.

4.8. Inertial Motion. The surface φ (U, S, V) = 0 corresponds to the all equilibrium states of the medium. The isolated medium spontaneously moves to the thermodynamic equilibrium state. It reaches his at any point of this surface. The internal forces will bring the medium to equilibrium if the entropy will be maximized and the internal energy will be minimized [3 - 5].

This work supposes that the impulse of the internal forces is arisen in the medium during a time period of its motion to the equilibrium; this pulse in conserved under compensation of external forces or their absence. In this case instead of the rest state the inertial motion of the medium without work production is possible. The inertial motion of the medium is described by the trajectory in the surface φ (U, S, V) = 0 which can intercept the regions of positive, negative or zero pressure in a general case.

4.9. Homogeneous and Heterogeneous Phase Equilibrium. The Gibbs' conditions for equilibrium and stability of thermodynamic media with a sign-alternate pressure found out in this work are allowing building two rules:
1) The equilibrium phases are stable to the segregation under inertial motion or in rest if they have pressure of opposite signs.
2) The equilibrium phases exist separately under the inertial motion or in rest if they have a pressure of the same sign.

Fest rule correspond to the homogeneous phase equilibrium or unusual medium. Second rule correspond to the heterogeneous phase equilibrium or usual medium.

## 5. ZERO PRESSURE AND PHASE EQUILIBRIUM

Zero pressure does not mean zero temperature. The

temperature corresponding to zero pressure, can be found with function

$$\omega \equiv p/u = \alpha T - 1, \quad (9)$$

that is obtained by substitutions $u \equiv U/V$, $s \equiv S/V$ and $\alpha \equiv s/u$ into (1). It has a useful for us property that for any positive number $\alpha$ we can find a zero value of $\omega$ and positive temperature $T_\alpha = 1/\alpha$. Consequently the value $\omega$ at $p = 0$ and $T = 1/\alpha$ divides the surface $\varphi(U, S, V) = 0$ into two regions: one with positive pressure at $T > T_\alpha$ and one with negative pressure at $T < T_\alpha$.

The isolated medium moves with time into the thermodynamic equilibrium state and reaches it at the temperature $T_0$ and positive pressure. The medium cannot leave the rest state by itself. Under unusual condition of the isolation when external forces are absent the primary medium can save the beginning mechanical impulse of the quasi-static expansion. Then the equilibrium medium will be cooled adiabatically, expanding by the inertia. In our case the medium is isolated, $U_0$ and $S_0$ are not variable, $\alpha$=const and the function (9) is linear.

The beginning state of inertial motion of the primary medium is presented by the point O on the Figure 1. In the geometrical interpretation of the medium inertial motion we use the fact that the identical parameter $\omega = -1$ exists for all functions $\omega = \omega(T)$ at zero temperature. According to equation (9), the primary medium has the phase stability under expansion along curve $oa$ under scenario 1i until the temperature $T_a$.

At the point $a$ the medium decay on the true new phases is possible due to the scenario 1i or 2i or 1j. The scenarios 1i and 2i describe the usual medium; the unusual medium appears under scenario 1j.

Let's realize scenario 1j. When primary medium decay is finished, the true new phases have opposite pressure signs at the temperature $T_a$ in the equilibrium state 1 and 2. If $p_1 = -p_2$, then $\omega_1 = -\omega_2$ at the elementary condition $u_1 = u_2$. Then segment $b1$ is part of function

$$\omega_1 = \alpha_1 T - 1, \quad (10)$$

the segment $b2$ is part of function

$$\omega_2 = -\alpha_1 T + 1. \quad (11)$$

These segments are the coexisting curves for phase 1 and 2. It is obvious that the equilibrium of phase 1 and 2 c with alternating-sign pressure is possible between the point $b$ and $a$ or in the temperature interval $\{T_b, T_a\}$.

In point $b$ phase 1 and 2 c with alternating-sign pressure achieve the zero pressure and at the temperature $T_b < T_a$ they spontaneously decay on phases 3 and 4 under the scenario 1j. The equilibrium of the phase 3 and 4 with alternating-sign pressure is possible in the temperature interval $\{T_c, T_b\}$.

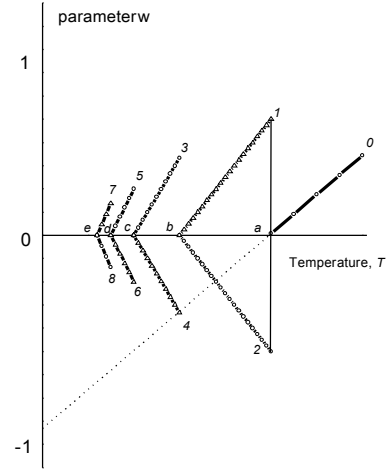

Figure 1: Illustration of function (9) and the inertial medium motion with and without alternating-sign pressure. Curve ao is beginning stage of usual medium motion. Points a, b, c, d and e are zero pressure the states. Isotherms 12, 34, 56 and 78 represent the decay of zero pressure medium on alternating-sign pressure phases.

In point $c$ at $T = T_c$ distinction of properties of phase 3 and 4 disappears and a zero pressure medium is generated which spontaneously decays on phases 5, 6 with alternating-sign pressure.

Further at $T = T_d$ and $\omega = 0$ spontaneous decay of zero pressure medium on phase 7 and 8 with alternating-sign pressure is possible and so on. This process can be called a fractal-like multiphase recombination or the process by which the equilibrium phases with alternating-sign pressure achieve to zero pressure again.

It is important that radiation cannot be completely transformed into condensate. The work must be done. However, the work is not product by the recombination cycle a1b2a or other one.

The medium can begin the inertial motion in the point 2 on the Figure 1. In this case the same fractal-like multiphase recombination takes place in further according to scenario 3j and the presented above scheme with the non-zero compensation.

## 6. ZERO ISOBAR AND PHASE EQUILIBRIUM

According to Figure 1 and the conditions (5), for the homogeneous phase equilibrium the equalities

$$\omega_1 + \omega_2 = 0 \text{ at the } u_1 = u_2,$$
$$\omega_3 + \omega_4 = 0 \text{ at the } u_3 = u_4,$$
$$\omega_5 + \omega_6 = 0 \text{ at the } u_5 = u_6,$$
$$\omega_7 + \omega_8 = 0 \text{ at the } u_7 = u_8$$

are valid and the curve $\omega = 0$ is the zero isobar too. The curves on the Figure 1 are the inertial motion trajectories of the phases 1 and 2 in the surface φ (U, S, V) = 0. In general case, $p_1 = -p_2$ at the conditions $\omega_1 \neq -\omega_2$ and $u_1 \neq u_2$. Then the inertial motion trajectories of the phases 1 and 2 won't be straight lines on the Figure 1. The role playing by zero isobar under homogeneous phase equilibrium is not changed.

A zero isobar from the scenarios i-j is a boundary curve between the homogeneous and heterogeneous phase equilibrium regions on the surface φ (U, S, V) = 0. So, according to the Gibbs' stability criterion, any medium with zero pressure coexists with zero pressure phases (scenario 1i). It is unstable relative to new phases with positive pressure (scenario 2i) and is stable relative to the new phases with negative pressure (scenario 3i). The new phases and initial medium are segregated.

Without segregation, the zero pressure medium coexists with phases which have opposite pressure signs (scenario 1j). This results in the zero pressure curve dividing the surface φ(U, S, V)=0 into two parts which stand for the homogeneous states of the primary medium with positive pressure (scenario 2j) and other homogeneous states of secondary medium consisting of the new phases with alternating sign pressure which are not primary medium (scenario 3j). If the first part of the surface is given, the other part can be calculated.

Corresponding to (1), at the each point of the zero isobar the product of temperature and $U/S \equiv 1/\alpha$ equals to unity. According to the rules 1 and 2, the medium under inertial motion intercepts the zero isobar and excludes the phase segregation by itself when $T/\alpha \leq 1$.

## 7. INERTIAL MOTION AND PHASE RULE

According to (1), the inertial motion of the medium takes place above the zero isobar if U<TS. For the zero isobar U=TS and the spontaneous phase transition is possible corresponding to scenarios 1i or 1j. The inertial motion of the medium takes place below the zero isobar corresponding to scenario 3j.

Regarding (1) and Gibbs' phase rule, the one-component system in the heterogeneous equilibrium may be realized no more, than from three phases. If the phases number reaches its maximum the inertial motion of the medium finishes due to absence of the thermodynamic degrees of freedom. According to scenarios 1i and Gibbs' phase rule, the inertial motion of the medium with positive pressure stops in triple point.

There is no evidence that the Gibbs' phase rule is not valid for homogeneous phase equilibrium. According to (3), the phases have different pressure and Gibbs had pointed out the constant mass ratio as a peculiarity of these phases [1]. It is also kept for an inertial motion. Regarding scenarios j and Gibbs' phase rule, the one-component system in the homogeneous equilibrium has one state with zero thermodynamic degrees of freedom. According to scenario 1j, three phases coexist under zero pressure only. So zero isobar is two phase boundary curve with other triple point for one-component system. The inertial motion finishes at this point.

## 8. THERMAL RADIATION AND ITS CONDENSATE AS MEDIUM CLAIMER S ON HOMOGENEOUS PHASE EQUILIBRIUM

8.1. Thermal Radiation as a Perl of Thermodynamics. Thermodynamics of thermal radiation is a cradle of the quant theory. Thermal radiation is a unique thermodynamic system while the expression

$$dU=TdS-pdV$$

for internal energy U, entropy S, and volume V holds the properties of the fundamental equation of thermodynamics regardless the variation of the photon number [3, 5]. Differential expression

$$dp/dT=S/V$$

for pressure p and temperature T is valid for one-component system under phase equilibrium if the pressure does not depend on the volume V [4,5]. Thermal radiation satisfies these conditions but shows no phase equilibrium. However, successful attempts of finding thermal radiation condensate in any form are unknown too. This work aims to support enthusiasm of experimental physicists and reports for the first time the phenomenological study of the thermodynamic medium consisting of thermal radiation and its condensate.

8.2. The Hypotheses about Condensate of Radiation. It is known [3, 5], that evolution of radiation or its inertial motion is impossible without participating matter and it realizes with absorption, emission and scattering of the beams as well as with the gravitational interaction. Transfer of radiation and electron plasma to the equilibrium state is described by the kinetic equation. Some of its solutions are treated as effect of accumulation in low-frequency spectrum of radiation, as Bose-condensation or non-degenerated state of radiation [6-10]. A known hypothesis about Bose-condensation of relic radiation and condensate evaporation has a condition: the rest mass of photon is thought to be non-zero [11]. Nevertheless, experiments show that photons have no rest mass [12].

8.3. Thermal Radiation Stability. The determinant of the stability [4] of equilibrium radiation

$$D = (\partial^2 U/\partial^2 S)_V \, (\partial^2 U/\partial^2 V)_S - \partial^2 U /(\partial S / \partial V)$$

is equal to zero because [5]

$$U(S, V) = \sigma V(3S/4\sigma V)^{4/3}.$$

While the zero determinants are related to the limit of stability, there are no thermodynamic restrictions for phase equilibrium of thermal radiation [4].

8.4. Thermal Radiation Condensate as Usual Medium. Radiation, matter and condensate may form a total thermal equilibrium. According to the transitivity principle of thermodynamic equilibrium [3, 5], participating condensate does not destroy the equilibrium between radiation and matter. Suppose that matter is a thermostat for the medium consisting of radiation and condensate. A general condition of thermodynamic equilibrium is an equality to zero of virtual entropy changes $\delta S$ or virtual changes of the internal energy $\delta U$ for media [4, 5]. Using indices for describing its properties, we write

$$S = S_{rad} + S_{cond},$$

$$U = U_{rad} + U_{cond}.$$

The equilibrium conditions

$$\delta S_{rad} + \delta S_{cond} = 0,$$

$$\delta U_{rad} + \delta U_{cond} = 0$$

will be completed by the expression

$$T\delta S = \delta U + p\delta V,$$

and then we get an equation

$$(1/T_{cond} - 1/T_{rad})\delta U_{cond} +$$
$$(p_{cond}/T_{cond})\delta V_{cond} + (p_{rad}/T_{rad})\delta V_{rad} = 0. \quad (12)$$

If $V_{rad} + V_{cond} = V = const$ and $\delta V_{rad} = -\delta V_{cond}$, then for any values of variations $\delta U_{cond}$ and $\delta V_{cond}$ we find

$$T_{rad} = T_{cond} = T,$$

$$p_{rad} = p_{cond} = p$$

as the equilibrium conditions. According to scenario 1i, condensate and radiation are segregated.

8.5. Thermal Radiation Condensate as Unusual Medium. When $V_{rad} = V_{cond} = V$ and $\delta V_{rad} = \delta V_{cond}$, the equilibrium conditions

$$T_{rad} = T_{cond}, \qquad p_{rad} = -p_{cond} \quad (13)$$

are satisfied. According to scenarios j, condensate and radiation are not segregated. Condensate is absolutely transparent for radiation; the radiation is integrated into the condensate. The positive and negative phase pressure means that the energy of radiation and condensate has limit. Then thermal radiation and condensate together are an unusual homogeneous medium.

8.6. The Entropy of Thermal Radiation and Condensate in Equilibrium. Let's consider the evolution of the condensate being in equilibrium with radiation. Once the medium is appeared, this medium consisting of the equilibrium condensate and radiation can continue the inertial adiabatic extension due to the assumed absence of external forces.

Figure 2 plots a curve of radiation extension as a cubic parabola

$$s_{rad} = 4\sigma T^3/3, \quad (14)$$

where $\sigma$ is the Stefan-Boltzmann constant, $S_c$ – constant from the third law of thermodynamics. This value is unknown. It is common that $S_c = 0$ at $T = 0$. In this work $S_c$ is a non-zero constant at any T.

Despite the fact that the entropy density $s_{cond}$ of condensate is unknown, we can show it in the Figure 2 as a set of positive numbers $\lambda = Ts$, if each $\lambda_i$ is ascribed an equilateral hyperbola

$$s_{cond} = \lambda_i/T.$$

Figure 2 illustrates both curves.

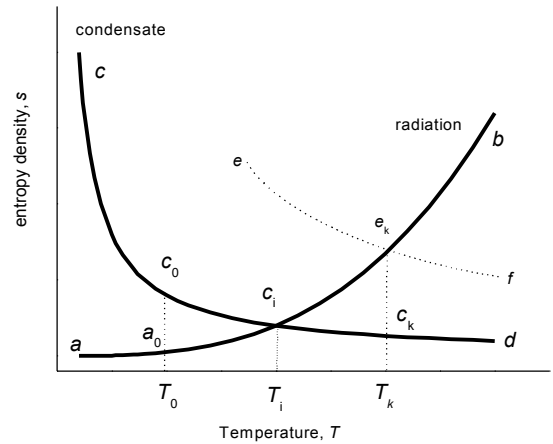

Figure 2: plots the density of entropy for radiation (curve ab) and for a condensate (curves cd and ef) are schematically plotted. The positive pressure of the radiation and negative pressure of condensate are equal by absolute value at the points $c_i$ and $e_k$ at the interceptions of these curves.

We include the cross-section point $c_i$ of the hyperbola cd and the cubic parabola ab in the Figure 2 in the interval $[c_0, c_k]$. Assume the entropy generation along the line cd outside this interval and the limits of the interval are fixing the boundary of the medium stability.

Absence of the entropy generation inside the interval $[c_0, c_k]$ means that the product $2s_i(T_k - T_0)$ is

$$_{T_0}\!\!\int^{T_k} dT\,(s_{cond} + s_{rad}).$$

By substituting s we can see that these equalities are valid only at $T_0=T_k=T_i$. So, if the condensate and radiation are in equilibrium, the equality

$$s_{cond} = s_{rad} = 4\sigma T^3/3 \qquad (15)$$

is also valid. Thus, when the equilibrium state is achieved the medium extension is realized along the cross-section line of the parabola and hyperbolas. Equalities (15) are of fundamental character; all other thermodynamic values for the condensate can be derived from these expressions.

8.7. Thermal Radiation with Positive Pressure. In the classical electromagnetisms the internal energy density of thermal radiation is

$$u_{rad} = 3p_{rad}, \qquad (16)$$

In other site, the internal energy density u(T) of thermal radiation is equal to

$$u(T)_{rad} = \int_0^\infty u(\nu, T)\, d\nu = \sigma T^4, \qquad (17)$$

where u(ν, T) is the function of the internal energy density depending on frequency ν and temperature T of the radiation. The agreement of theoretical distribution and experimentally observed one u(ν, T) is found by Planck in framework of the quantum hypothesis [3, 5]. The pressure and the entropy density $s_{rad}$ are

$$s_{rad} = 4\sigma T^3/3, \qquad p_{rad} = \sigma T^4/3. \qquad (18)$$

The second law of thermodynamics includes the value u too. It can be written as

$$u = Ts - p, \qquad (19)$$

where u and s are densities of energy and entropy, respectively. Substituting the values $u_{rad}$, $s_{rad}$ and $p_{rad}$ from (16)-(18) in the equation (19), one can conclude that the experiment with an ideal radiator does not contradict to the classical electromagnetisms, quant theory and thermodynamics of thermal radiation.

8.8. Radiation with Zero and Negative Pressure. The radiator with zero or negative pressure of radiation is unknown. The temperature of radiation in these states is positive and we assume that internal energy of radiation is equal to $\sigma T^4$. According to (9), (19), the internal energy of radiation with entropy $S_0$ at zero pressure and positive temperature $T_\alpha$ is excessive:

$$u_{p=0} = Ts_0 - 0 = \sigma T^4 + \sigma T^4/3.$$

The radiation should be received energy $\sigma T^4/3$ from the environment at the reaching a zero isobar along a surface φ (U, S, V) = 0.

If the initial medium with entropy $S_0$ reaches zero pressure not along a surface φ (U, S, V) = 0 during the adiabatic process, the energy difference is equal to

$$u_{p=0} - (Ts_0 - p_{rad}) = 4\sigma T^4/3 - 4\sigma T^4/3 + 0 = 0.$$

In this case the radiation should be not received energy from the environment.

Assume that the primary radiation moves along the curve *a*(-1) at the negative pressure. In this case

$$p^*_{rad} = -\sigma T^4/3$$

and the internal energy will be

$$u^*_{rad} = Ts^* - p^*_{rad} = 5\sigma T^4/3. \qquad (20)$$

It should be emphasized that the value $u^*_{rad}$ is the internal energy radiation with the negative pressure and radiation is out of the equilibrium with condensate. In this case $u^*_{rad} > u(T)_{rad}$. The excessive energy of radiation is equal to

$$\Delta u^* = u^*_{rad} - u_{rad} = 2\sigma T^4/3$$

and should be received from the environment for crossing of a zero isobar and cooling more low temperature $T_a$.

Thus, radiation with zero and negative pressure has excessive energy. These states are not reached during the inertial move thermal radiation.

8.9. Equilibrium between Thermal Radiation and Condensate. Radiation with negative pressure is instability itself and performs the phase transformation forming the homogeneous equilibrium between radiation with positive pressure and condensate with negative pressure under scenarios i-j. According to the equilibrium conditions (5), (13) and (18), the condensate pressure is

$$p_{cond} = -\sigma T^4/3.$$

This phase transformation carries out at the condition $S_0 = S_{rad} + S_{cond}$. As $p_{rad} = -p_{cond}$, then

$$u_{rad} + u_{cond} = Ts_0.$$

According to (16) – (18),

$$u_{cond} = 4\sigma T^4/3 - \sigma T^4 = \sigma T^4/3 = -p_{cond}.$$

Thus, for the homogeneous phase equilibrium between thermal radiation and condensate the equalities

$$\omega_1 + \omega_2 \neq 0 \text{ at the } u_1 = 3u_2,$$
$$\omega_3 + \omega_4 \neq 0 \text{ at the } u_3 = 3u_4,$$
$$\omega_5 + \omega_6 \neq 0 \text{ at the } u_5 = 3u_6,$$
$$\omega_7 + \omega_8 \neq 0 \text{ at the } u_7 = 3u_8$$

are valid but the curve ω = 0 remains the zero isobar.

Assume that the non-equilibrium initial medium with entropy $S_0$ reaches the surface φ (U, S, V) = 0 in the point with negative pressure. It decays spontaneously according to the scenario 3j. According to (14), (15) and (20), the energy

$$u^* - (u_{rad} + u_{cond}) = 5\sigma T^4/3 - 4\sigma T^4/3 = \sigma T^4/3$$

is uncompensated energy which evolve during the phase transformation under the scenario 3*j* for the instability primary radiation at the negative pressure and the medium consisting of thermal radiation and condensate. The uncompensated energy can be dissipated in space.

*8.10. Thermal Radiation and Condensate out of Equilibrium.* The extension of radiation-condensate medium is an inerial process, so that the positive pressure of radiation $p_{rad}$ lowers, and the negative pressure of the condensate $p_{cond}$ increases according to the condition (13).

Matter is extended with thermal radiation. As it is known in cosmological theory [3, 5], the plasma inertial extension had led to formation of atoms and distortion of the radiation-matter equilibrium. Further local matter non-homogeneities were appeared as origins of additional radiation and, consequently, matter created a radiation excess in the medium after the equilibrium radiation-matter was disturbed. This work supposes that radiation excess may cause equilibrium displacement for the medium, thus radiation and condensate will continue inertial extending in different equilibrium processes in parallel.

We assume that the distortion of the equilibrium radiation-condensate had been occurred at the temperature $T_i$ of the medium at the point $c_i$ in the Figure 2. The radiation will be extended adiabatically along the line $c_i a$ of the cubic parabola without entropy generation. While the condition $V_{rad}=V_{cond}$ is satisfied if the equilibrium is disturbed, the equality $s_{cond}=s_{rad}$ points out directions of the condensate extension without entropy generation. As it is shown in the Figure 2, the unchangeable adiabatic isolation is possible if the condensate extends along the isotherm $T_i$ without heat exchange with radiation. Differentiation of the expression

$$U_{cond} - T_i S_{cond} + p_{cond}V = 0$$

with $T$=const and $S$=const gives that $p_{cond}$ is also constant.

The medium as a whole extends in such a manner that the positive pressure $p_{rad}$ of radiation decreases, and the negative pressure $p^*_{cond}$ remains constant. As radiation cools down, the dominant $p^*_{cond}$ of the negative pressure arises, and the medium begins to extend with positive acceleration.

## 9. CONCLUSION

As a conclusion one should note that the negative pressure of the condensate of thermal radiation is Pascal-like and isotropic, it is constant from the moment as the equilibrium with radiation was disturbed by the condensate and is equal (by absolute value) to the energy density. The condensate of thermal radiation is a physical medium which interacts only with the radiation and this physical medium penetrates the space as a whole. This physical medium cannot be obtained under laboratory conditions because there are always external forces for a thermodynamic system in laboratory. As known [13], the photon Bose-condensate is obtained. This condensate has no negative pressure while it is localized in space. It seems very interesting to find in the nature a condensate of thermal radiation with negative pressure. Possible forms of physical medium with negative pressure and their appearance at cosmological observations are widely discussed [14]. The radiation can consist of other particles, and then the photon, among them may be also unknown particles. We hope that modeling the medium from the condensate and radiation will be useful for checking the hypotheses and will allow explaining the nature of the substance responsible for accelerated extension of the Universe. The medium from thermal radiation and condensate is the first indication of the existence of physical vacuum as one of the subjects in classical thermodynamics and the complicated structure of the dark energy [15].

## References

[1]. J.W. Gibbs, *The Scientific Paper of J.W. Gibbs, Thermodynamics*, vol. 1, Kessinger Publishing, New York, 2007.

[2]. C. Kittel and H. Kroemer, *Thermal Physics,* Freeman, New York, 2000.

[3]. I. Prigogin and D. Kondepudi, *Modern Thermodynamics*, John Wiley & Sons, New York, 1999.

[4]. A. Muenster, *Classical Thermodynamics*, Wiley-Interscience, New York, 1971.

[5]. I.P. Bazarov, *Thermodynamics*. Pergamon Press, Oxford, 1991.

[6]. A.S. Kompaneets, "The establishment of thermal equilibrium between quanta and electrons"*, Soviet Physics - JETP*, Vol. 4, no. 5, pp. 730-740, 1957.

[7]. H. Dreicer, "Kinetic Theory of an Electron‐Photon Gas", *Phys. Fluids*, vol. 7, no. 5, pp. 732-754, 1964.

[8]. R. Weymann, "Diffusion Approximation for a Photon Gas Interacting with a Plasma via the Compton Effect ", *Phys. Fluids*, vol. 8, no. 11, pp. 2112-2114, 1965.

[9]. Ya.B. Zel'dovich and R.A. Syunyaev, "Shock wave structure in the radiation spectrum during bose condensation of photons", *Soviet Physics - JETP*, vol. 35, no. 1, pp. 81-85, 1972.

[10]. A.E. Dubinov, "Exact stationary solution of the Kompaneets kinetic equation", *Technical Physics Letters*, vol.35, no.3, pp.260-262, 2009.

[11]. V.A. Kuz'min and M.E. Shaposhnikov, "Condensation of photons in the hot universe and longitudinal relict radiation", *JETP Lett.*, vol. 27, no. 11, pp.628-631, 1978.

[12]. G. Spavieri and M. Rodrigues, "Photon mass and quantum effects of the Aharonov-Bohm type", *Phys. Rev. A*, vol. 75, 05211, 2007.

[13]. J. Klaers, J. Schmitt, F. Vewinger and M. Weitz, "Bose–Einstein condensation of photons in an optical microcavity", *Nature*, vol. 468, pp. 545-548, 2009.
[14]. P. J. E. Peebles and Bharat Ratra, "The cosmological constant and dark energy", Reviews of Modern Physics, vol. **75** (2), pp. 559–606, 2003.
[15]. V.I. Laptev and H. Khlyap, "Photons as Working Body of Solar Engines", in *Solar Cells: new aspects and solution*, Edited by L.A. Kosyachenko, pp. 357-396, INTECH, 2011.